\documentclass[11pt]{article}
\usepackage{a4,parskip,float}
\usepackage{epsfig}
\newcommand{\om}{$\omega$ meson }
\newcommand{\np}{\\[0.5cm]}
\begin{document}
\begin{titlepage}
\vspace*{-2cm}
\begin{flushright}
\bf
TUM/T39-01-07
\end{flushright}
\vskip 2.5cm
\begin{center}
{\LARGE\bf The $\omega$ meson at high temperatures\footnote{Work supported in part by BMBF and GSI.}\\}
\vspace{1.5cm}
{\large R. A. Schneider\footnote{e-mail:{\tt schneidr@ph.tum.de}} and W. Weise\footnote{e-mail:{\tt weise@ect.it}} \\}
\date{\today{}}
\vspace{1.5cm}
{\it Physik-Department, Theoretische Physik\\ Technische Universit\"at M\"unchen, D-85747 Garching, Germany \\[0.5cm] and \\[0.5cm]
ECT$^*$\\ I-38050 Villazzano (Trento), Italy\\}
\vspace*{3cm}
{\bf Abstract\\}
\bigskip
\begin{minipage}{15cm}
The decay of the $\omega$ meson in a heat bath of thermally excited pions is studied within the framework of real-time thermal field theory using an appropriate effective Lagrangian. We show that the $\omega$ meson spectrum broadens considerably at temperatures $T > 100$ MeV, primarily because of $\omega\pi \rightarrow \pi\pi$ reactions in the thermal environment.
\end{minipage}
\end{center}
\end{titlepage}
\newpage
The hadronic phase of QCD presumably undergoes a transition towards chiral restoration and deconfinement at temperatures around $T_C \simeq 150 - 160$ MeV, according to lattice QCD thermodynamics with three quark flavours \cite{FK}. Changes of vector meson spectra in a thermally excited hadronic environment are discussed as possible pre-signatures for such a transition. In this paper we investigate the spectral function of the \om as it evolves with increasing temperature $T$. Our tool is thermal field theory based on an effective Lagrangian of interacting pions and vector mesons. At high temperatures, the primary decay $\omega \rightarrow 3\pi$ should be modified by the presence of thermally excited pions. In addition, reactions such as $\omega\pi \rightarrow \pi\pi$ in the pionic heat bath are now possible. The resulting thermal broadening of the \om spectrum should be of some relevance to $e^+e^-$ production in ultrarelativistic heavy-ion collisions at the CERN-SPS and at RHIC.
\vskip 0.5cm
The primary decay mode of the \om  in vacuum is $\omega \rightarrow \pi^+\pi^0\pi^-$ with a branching ratio of 89\% and a width $\Gamma_{\omega \rightarrow 3\pi} \simeq $ 7.6 MeV. The $\omega$ coupling to pions involves two parts: an anomalous direct $\omega \leftrightarrow 3\pi$ interaction and a two-step interaction $\omega \leftrightarrow \rho\pi \leftrightarrow 3\pi$ (the Gell-Mann, Sharp, Wagner (GSW) process \cite{GSW}) in which the second step includes the $\rho \rightarrow \pi\pi$ decay. The effective Lagrangian describing the decay dynamics is taken from ref.\cite{KKW}:
\begin{eqnarray}
\mathcal{L}_{int} & = & \frac{G}{f_\pi} \ \epsilon^{\mu\nu\alpha\beta} \partial_\mu \omega_\nu (\partial_\alpha \rho_\beta^- \pi^+ + \partial_\alpha \rho_\beta^0 \pi^0 + \partial_\alpha \rho_\beta^+ \pi^-) \nonumber \\ & & + \ i \frac{H}{f_\pi^3} \ \epsilon^{\mu\nu\alpha\beta} \omega_\mu \partial_\nu \pi^+ \partial_\alpha \pi^0 \partial_\beta \pi^- \nonumber \\ & & - \ i g \ \rho^0_\mu (\pi^+ \partial^\mu \pi^- - \pi^- \partial^\mu \pi^+) + \dots \ , \label{Lag}
\end{eqnarray}
with the pion decay constant $f_\pi$ = 92.4 MeV. The constants $G = 1.2$ and $H = 0.18$ are determined by the branching between $\omega \rightarrow \rho\pi$ and the direct $\omega \rightarrow 3\pi$ channel, while $g \simeq 6$ reproduces the $\rho \rightarrow \pi\pi$ decay width. 
\np
The self-energy of the \om in the thermal environment is represented as  a tensor $\Pi_{\mu\nu}(E,\vec{p}; T)$ as a function of the $\omega$ four-momentum $p = (E, \vec{p})$ and the temperature $T$. The heat bath defines a distinguished frame of reference, and Lorentz invariance is broken. In the present paper we consider the \om at rest with respect to the heat bath, {\em i.e.} with $\vec{p} = 0$. In this case the longitudinal and transverse parts, $\Pi_L$ and $\Pi_T$, of the \om self-energy coincide, and there is a single scalar function,
\begin{equation}
\Pi(E,T) \equiv - \frac{1}{3} \Pi_{\ \mu}^\mu (E, \vec{p} = 0; T),\label{selfenergy}
\end{equation}
which summarizes the complete information about the in-medium interactions of the $\omega$ meson.
\np
We use real-time thermal field theory \cite{LB} to derive the temperature dependent \om spectral function. Within this framework the thermal \om propagator,
\begin{equation}
{\mathbf D} = {\mathbf D}_F + {\mathbf D}_F (-i{\mathbf \Pi}) {\mathbf D},
\end{equation}
as well as the self-energy ${\mathbf \Pi}$ are represented in the form of 2 x 2 matrices. Diagonalisation of ${\mathbf D}$ yields the self-energy function $\Pi$ of eq.(\ref{selfenergy}) which describes the thermal decays $\omega \rightarrow \pi\pi\pi$ as well as $\omega \pi \rightarrow \pi\pi$. 
\np
Let the pion four-momenta be denoted by $q_i = (E_i, \vec{q}_i)$ with $i = 1,2,3$. Then the imaginary part of $\Pi$ becomes \footnote{This relation is actually valid for $n$-particle decays when the upper limit 3 in sums and products is replaced by $n$.}
\begin{eqnarray}
& & \mbox{Im}\Pi(p)  =  \frac{1}{2} \left( \prod\limits_{i=1}^{3} \int \frac{d^3 q_i}{2 E_i (2\pi)^3} \right) (2\pi)^4 \times \nonumber \\ & \times & \left\{ \delta^{(4)}(p - \sum_i q_i) \ | \mathcal{M}(p \rightarrow \sum_i q_i) |^2 \left( \prod\limits_{i=1}^{3} (1 + n_i) - \prod\limits_{i=1}^{3} n_i \right)  + \ \dots \ \right\}, \label{master}
\end{eqnarray}  
where the dots indicate all permutations with $q_i \rightarrow -q_i$ and corresponding re-arrangements of the Bose factors 
$$
n_i \equiv f_B(E_i) = \frac{1}{e^{\beta E_i} -1},
$$
with $\beta = 1/T$. Explicitly,
\begin{eqnarray}
\mbox{Im}\Pi(p) & = & \frac{1}{8 (2\pi)^5} \int \frac{d^3 q_1}{E_1} \int \frac{d^3 q_2}{E_2} \int \frac{d^3 q_3}{E_3} \nonumber \times \\ & \times & \left\{ \delta^{(4)}(p - q_1 - q_2 - q_3) \ | \mathcal{M}(p,q_1,q_2,q_3) |^2 \Big( (1 + n_1)(1 + n_2) (1 + n_3) - n_1 n_2 n_3) \Big) \right. \nonumber \\ & &  + \ \delta^{(4)}(p - q_1 - q_2 + q_3) \ | \mathcal{M}(p,q_1,q_2,-q_3) |^2 \Big( (1 + n_1)(1 + n_2) n_3 - n_1 n_2 (1 + n_3) \Big) \nonumber \\ & & + \ \mbox{permutations} \ \bigg\}, 
\end{eqnarray}
which includes the $\omega \rightarrow \pi_1\pi_2\pi_3$ channel together with the $\omega \pi_3 \rightarrow \pi_1\pi_2$ reaction, and permutations thereof, in the heat bath of thermally excited pions. 
\np
The matrix element $\mathcal{M}$ is calculated using the interaction Lagrangian (\ref{Lag}):
\begin{equation}
\mathcal{M}(p, q_1, q_2, q_3) = \frac{-H}{f_\pi^3} + \frac{2 G g}{f_\pi}\ \sum_{i = 1}^3 \frac{1}{(p - q_i)^2 - m^2_\rho -
\Pi_\rho(p - q_i)}.\label{M}
\end{equation}
The second term involves $\pi\rho$ intermediate states, with $m_\rho = 770$ MeV, and the zero temperature self-energy $\Pi_\rho$ given explicitly in ref.\cite{KKW} where a detailed discussion of the matrix element (\ref{M}) can be found. Note that through the replacement $q_i \rightarrow -q_i$ for one of the pion momenta, the process $\omega \pi \rightarrow \rho \rightarrow \pi\pi$ is included. We have assumed that effects of temperature on intermediate $\rho$ mesons are small (suppressed as $e^{-m_\rho/T}$) in the temperature range $T \leq 150$ MeV that we are interested in. 
\np
The thermal \om width is
$$
\Gamma_\omega(T) = \frac{\mbox{Im}\Pi(E = m_\omega, \vec{p} = 0; T)}{m_\omega}  
$$
at the physical $\omega$ mass $m_\omega$ which is now also temperature-dependent. The explicit calculation for the \om at rest ($p = (m_\omega, 0,0,0)$) leads to the result
$$
\Gamma_\omega   =   \frac{m_\omega}{192 \pi^3} \left( \mathcal{B}_1 + 3 \ \mathcal{B}_2 \right) ,
$$
with
\begin{eqnarray}
\mathcal{B}_1 & = & \int \limits_{\Delta_1} dE_+ dE_- \left( \vec{q}_+^{\ 2} \vec{q}_-^{\ 2} -
(\vec{q}_+ \cdot \vec{q}_-)^2 \right)|\mathcal{M}(p, q_+, q_0, q_-)|^2 \cdot n(E_+, E_-,
m_\omega), \nonumber \\ \mathcal{B}_2 & = & \int \limits_{\Delta_2} dE_+ dE_- \left( \vec{q}_+^{\ 2}
\vec{q}_-^{\ 2} - (\vec{q}_+ \cdot \vec{q}_-)^2 \right)|\mathcal{M}(p, q_+, -q_0, q_-)|^2 \cdot
(-1) n(E_+, E_-, m_\omega). \label{B1and2}
\end{eqnarray}
Here, $E_\pm$ and $\vec{q}_\pm$ are the energies and momenta of the two charged pions in the
final state. To keep our notation short, we have introduced the function
\begin{eqnarray*}
n(E_+, E_-, m_\omega) & = &  [1 + f_B(E_+)] \ [1 + f_B(E_-)] \ [1 + f_B(m_\omega -
E_+ - E_-)] \\ & & - f_B(E_+) \ f_B(E_-) \ f_B(m_\omega - E_+ - E_-).
\end{eqnarray*}
In eq.(\ref{B1and2}), $\mathcal{B}_1$ describes the process of a Bose-enhanced three-pion 
decay of the $\omega$ meson where $n(E_+, E_-, m_\omega)$ accounts for the characteristic Bose factors.
The kinematically allowed integration region in the $E_+ E_-$ plane, $\Delta_1$, is the same as in
the $T=0$ case for the decay process. It is limited to $m_\pi \le E_\pm \le m_\omega - 2m_\pi$ and
$2m_\pi \le E_+ + E_- \le m_\omega - m_\pi$ with the constraint $ \vec{q}_+^{\ 2} \vec{q}_-^{\ 2} -
(\vec{q}_+ \cdot \vec{q}_-)^2 > 0$.
\np
To interpret the term $3 \ \mathcal{B}_2$ in eq.(\ref{B1and2}), we note that
\begin{eqnarray}
 - n(E_+, E_-, \omega) & = &  [1 + f_B(E_+)] \ [1 + f_B(E_-)] \  f_B(E_+ + E_- - \omega) \nonumber \\ & &
- f_B(E_+) \ f_B(E_-) \ [1 + f_B(E_+ + E_- - \omega)]\label{minusn}
\end{eqnarray}
by using $1 + f_B(E) + f_B(-E) = 0$. Thus $\mathcal{B}_2$ corresponds to a scattering of
the $\omega$ meson off a thermally excited $\pi^0$ into $\pi^+$ and $\pi^-$, where each particle
has the characteristic thermal corrections. The integration region $\Delta_2$ is still constrained
by $ \vec{q}_+^{\ 2} \vec{q}_-^{\ 2} - (\vec{q}_+ \cdot \vec{q}_-)^2 > 0$, but now in the region defined by
$E_+ + E_- \ge m_\omega + m_\pi$.
Because the integration region $\Delta_2$ is now unbounded and
the functions defined on it are only damped by Bose factors and their combinations, the
scattering contribution to the decay width increases quite strongly with temperature, as we will
see below.
The factor 3 in front of $\mathcal{B}_2$ stems from the fact that the pion charges are merely labels to
distinguish the particles, and there are three possible combinations. Isospin symmetry implies that the contributions from $\omega + \pi^+ \rightarrow \pi^0 + \pi^+$ and $\omega + \pi^- \rightarrow \pi^0 +
\pi^-$ are the same as $\omega + \pi^0 \rightarrow \pi^- + \pi^+$ which we have chosen as our reference calculation. Note that processes such as $\omega\pi\pi \rightarrow \pi$ are formally present in Eq.(\ref{master}), but they do not contribute since they are kinematically forbidden.
\np
The real part of the decay diagram, {\em i.e.} the mass shift due to the 3-pion
loop, has not been calculated so far. It involves a two-dimensional Cauchy Principal Value
integral which is impossible to compute analytically and difficult to calculate
numerically. However, we expect the thermal mass shift of the \om to be small, recalling that for the $\rho$ meson, the leading thermal mass shift comes only at order $T^4$ \cite{DEI,E}. In the following we use $m_\omega = 783$ MeV.
\np
\begin{figure}[t]
\begin{center}
\epsfig{file=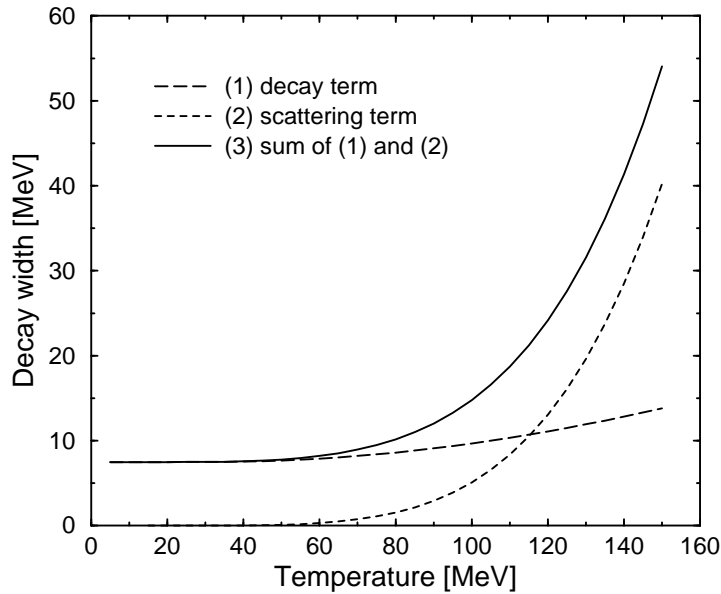,width=9cm, angle = -90} \caption[The omega decay width $\Gamma_\omega$
vs. temperature]{The omega decay width $\Gamma_\omega$ as a function of temperature. For
explanations see text. } \label{gamma_omega_plot}
\end{center}
\end{figure}
In Figure \ref{gamma_omega_plot} we plot the dependence of the $\omega$ meson decay width on
temperature. Shown are the contributions from the decay term $\mathcal{B}_1$ in eq.(\ref{B1and2}),
the scattering term $3 \ \mathcal{B}_2$ and the sum of both.
An interesting picture emerges: the direct $\omega \rightarrow 3\pi$ decay term behaves as we would expect from
our experience with 2-body decays. The Bose enhancement leads to a moderate rise of $\Gamma_\omega$
from its vacuum value 7.5 MeV up to about 14 MeV at $T = 150$ MeV. This increase is nevertheless stronger than in a
2-body decay because there are now three Bose enhancement factors present. The
contribution from the re-combination of thermal pions into an \om (which reduces the width) cancels only part of these
 enhancement factors.
\np
 At temperatures $ T > 60 $ MeV, a substantial
fraction of pions is excited, and the scattering term starts to play an important r\^ole. Its increase with temperature is, not surprisingly, reminiscent of the rate at which the thermal pion density grows. However, because of the additional Bose enhancement of the final state pions, the scattering rate rises stronger than $T^3$. In fact, a good fit over the temperature range considered is 
$$
\Gamma_{\omega\pi \rightarrow \pi\pi} = \left( \frac{T}{T_s} \right)^5 \mbox{ MeV} \quad \mbox{with} \quad T_s = 72.3 \mbox{ MeV}\simeq \frac{m_\pi}{2}.
$$
For $T > 120$ MeV, this width becomes larger than the pure decay term, rising up to about 40 MeV at $T = 150 $ MeV. The total decay
width is now the sum of decay and scattering terms, leading to $\Gamma_\omega \simeq 15$ MeV at $T=100 $ MeV and
$\Gamma_\omega \simeq 55$ MeV at $T=150$ MeV. At even higher temperatures one enters into the region of the chiral and deconfinement transition where the \om supposedly dissolves and releases its quark constituents.
\np
\begin{figure}[t]
\begin{center}
\epsfig{file=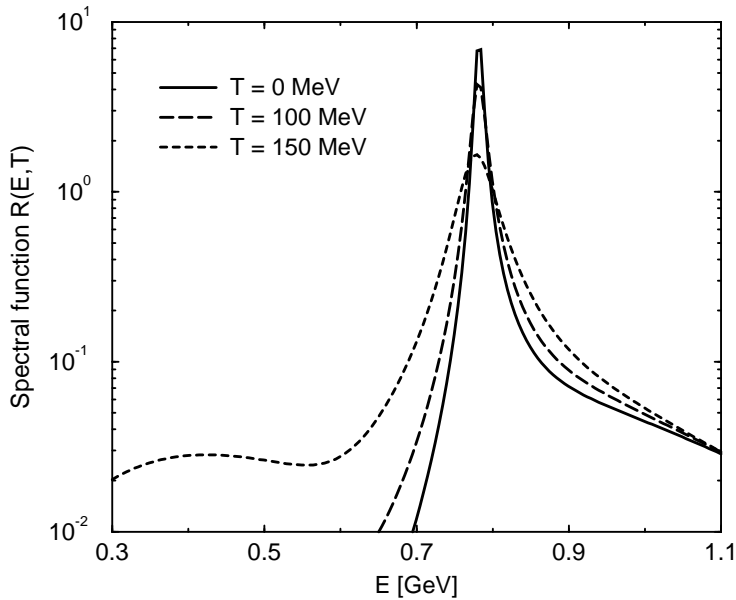,width=9cm, angle = -90} \caption{The spectral function
$R(E, T)$, defined in eq.(\ref{spectral}), for an \om at rest ($p =
(E, 0,0,0)$). Shown are the spectra for three different temperatures.}
\label{spectrum_omega_plot}
\end{center}
\end{figure}
In Figure \ref{spectrum_omega_plot} we show the (dimensionless) thermal electromagnetic spectral function,
\begin{equation}
R(E, T) = \frac{12\pi}{E^2} \ \mbox{Im} \bar{\Pi}(E,T), \label{spectral}
\end{equation}
expressed in terms of the (reduced) current-current correlation function $\bar{\Pi}$ in the $\omega$ channel. The imaginary part of the \om self-energy, Im$\Pi(E,T)$, is related to the correlator by \cite{KKW}
\begin{equation}
\mbox{Im} \bar{\Pi}(E) = \frac{\mbox{Im} \Pi(E, T)}{9 g^2} \ |F(E)|^2, \label{ImBarPi}
\quad \mbox{with} \quad F(E) = \frac{m_\omega^2}{E^2 - m_\omega^2 + i \mbox{Im}\Pi(E,T)}.
\end{equation}
Eq.(\ref{spectral}) is normalized such that it can be directly compared to the cross section rate $\sigma(e^+e^- \rightarrow \mbox{hadrons}) / \sigma (e^+ e^- \rightarrow \mu^+ \mu^-)$. We have ignored contributions from the $n$-pion continuum with $n \geq 5$, which add to the spectrum at $E > 1$ GeV but will not experience significant thermal effects. Hardly any change is visible in the spectral function at $T < 100$ MeV. However, as the temperature increases beyond 100 MeV the \om broadens substantially (the situation here is quite different from the one in matter at zero temperature and finite baryon density $\rho_B$ where the $\omega$ mass shift and the increase of its width have a leading linear dependence on $\rho_B$ \cite{KKW2}).    
\np
Different approximate calculations of the \om width \cite{ES,KH,ASRRS} lead to results smaller than ours (their values for $\Gamma_\omega$ lie between 20 and 40 MeV at $T = 150$ MeV; see however \cite{RP} where the simple estimate $\Gamma_\omega(T_C) \simeq 9 \Gamma_\omega(0)$ comes quite close to our value). Our result is a natural outcome of thermal field theory and does not require an {\em ad hoc} modification of a $T=0$ decay width formula. In fact, we suspect that the modified Breit-Wigner ansatz used in ref.\cite{ASRRS} is not appropriate at high temperatures where the $\omega \pi$ coupling to the isovector $\pi\pi$ continuum is strong and gives rise to the broad background in the low-mass spectrum, as shown in Figure \ref{spectrum_omega_plot}. Recently, a calculation of the \om width at finite temperature using forward scattering amplitudes inferred from experimental data yielded a width of 50 MeV at $T = 150$ MeV \cite{EBEK}, which supports a larger width than that obtained in refs.\cite{KH,ASRRS}.   
\np
We conclude that, at high temperatures approaching the critical range commonly associated with the chiral and deconfinement transition in QCD, the \om tends to loose its sharp resonance structure and experiences a strongly increased width, mainly as a consequence of collisions with thermal pions in the heat bath. The present results are of some relevance to the analysis of $e^+e^-$ rates produced in ultrarelativistic heavy ion collisions \cite{CERES, RW, SW}. While the \om visibly retains its quasi-particle structure even at high temperatures (see Figure \ref{spectrum_omega_plot}), its lifetime decreases to about 3 fm/$c$ at $T = 150 $ MeV. This implies that, contrary to common belief, the \om has a chance to interact and decay inside the expanding fireball produced in central heavy ion collisions.

\end{document}